\documentclass[prd,twocolumn,floatfix,nofootinbib,superscriptaddress]{revtex4}

\usepackage{amsmath}
\usepackage{amssymb}
\usepackage{bm}
\usepackage{epsfig}
\usepackage{graphicx}
\usepackage{longtable}
\usepackage{color}
\usepackage{nicefrac}
\usepackage{epstopdf}
\usepackage{tabularx}
\usepackage{dcolumn}  
\newcolumntype{L}[1]{>{\raggedright\arraybackslash}p{#1}}
\newcolumntype{C}[1]{>{\centering\arraybackslash}p{#1}}
\newcolumntype{R}[1]{>{\raggedleft\arraybackslash}p{#1}}

\usepackage[colorlinks=true, pdfstartview=FitV, linkcolor=blue, citecolor= red,urlcolor=blue]{hyperref}
\definecolor{darkgreen}{rgb}{0,0.5,0}
\definecolor{purple}{rgb}{0.5,0,0.5}
\definecolor{nblue}{rgb}{0.0,0.0,0.50}
\definecolor{scarlet}{rgb}{1.0,0.2,0}

\begin{document}

\title{Strong two-meson decays of light and charmed vector mesons}

\author{Roberto Correa da Silveira}
\affiliation{LFTC, Universidade Cidade de S\~ao Paulo, Rua Galv\~ao Bueno 868, S\~ao Paulo, SP 01506-000, Brazil}

\author{Fernando E. Serna}
\affiliation{LFTC, Universidade Cidade de S\~ao Paulo, Rua Galv\~ao Bueno 868, S\~ao Paulo, SP 01506-000, Brazil}
\affiliation{Departamento de F\'isica, Universidad de Sucre, Carrera 28 No.~5-267, Barrio Puerta Roja, Sincelejo 700001, Colombia}

\author{Bruno El-Bennich}
\affiliation{LFTC, Universidade Cidade de S\~ao Paulo, Rua Galv\~ao Bueno 868, S\~ao Paulo, SP 01506-000, Brazil}


\begin{abstract}
  We calculate the strong decay couplings for $\rho \to \pi\pi$, $\phi\to KK$, $K^* \to K\pi$ and $D^*\to D\pi$ in a unified and consistent approach based on the impulse 
  approximation, nonperturbative solutions of the quark-gap equation and the Poincar\'e invariant Bethe-Salpeter amplitudes of vector and pseudoscalar mesons.
  In particular, we obtain the coupling $g_{D^*\!D\pi} = 17.24^{+3.06}_{-2.30} $ in very good agreement with the experimental value by CLEO, which corresponds to 
  a strong effective coupling between heavy vector and pseudoscalar mesons to the pion of $\hat g = 0.58^{+0.10}_{-0.08}$.
\end{abstract}

\date{\today}
\maketitle


\section{Introduction}

One of the challenges in hadron physics is to understand the spectrum, constituent composition and momentum distribution of quarks and gluons within the hadrons. 
To obtain deeper insight into the hadron's structure, their excitations  have been intensively investigated in the past decades. This includes radial excitations, higher 
angular momentum states and exotic states containing constituent gluons that contribute to the total angular momentum of the hadron. 

The vector mesons, being the lowest spin excitations of the pseudoscalars, offer a first glimpse into an electromagnetic excitation of a $\bar qq$ pair. This is because neutral 
vector mesons can directly couple to the photon via an electromagnetic current since their quantum numbers, $J^{PC} = 1^{--}$, are those of the photon. Naturally, they have 
been much studied and from the viewpoint of functional approaches to Quantum Chromodynamics (QCD) they were helpful to establish the ladder truncation of the  
Bethe-Salpeter equation (BSE)~\cite{Maris:1999nt}, at least for the ground states of lighter vector mesons. Of course, beyond the masses of the pseudoscalar and vector 
mesons, their electromagnetic and electroweak properties are of fundamental interest and there is no lack of studies dedicated to weak decay constants, elastic and transition 
form factors~\cite{Maris:2000sk,deMelo:1997hh,Bakker:2002mt,deMelo:2012hj,DeMelo:2018bim,deMelo:2014gea,Chang:2013nia,Raya:2015gva,Ding:2018xwy,Xu:2019ilh,
daSilva:2012gf,El-Bennich:2012mkr,El-Bennich:2008dhc,Ivanov:2007cw,El-Bennich:2009gbu}.

Beyond these observables, the strong decays of vector mesons into two light(er) mesons provide another source of information on the nonperturbative dynamics
complementary to electromagnetic interactions and weak decays. They are the simplest possible decays that proceed via strong interactions and since the vector meson 
decays via a $P$-wave interaction, the Bethe-Salpeter amplitude (BSA) is probed differently than in the electroweak sector. In here, our main object of interest is the 
reaction $D^*\to D\pi$ which we studied in Refs.~\cite{El-Bennich:2010uqs,El-Bennich:2011tme,El-Bennich:2012hom,El-Bennich:2016bno} within the limitations of not 
having the BSA of charmed mesons at hand but motivated by the first measurement of the $D^*$ width, $\Gamma(D^{*+}) = 96\pm 4\pm 22$~keV~\cite{CLEO:2001sxb}. 
This result is of great interest, as it is one of the few quantities in flavor physics that does not probe electroweak properties of heavy mesons and which opens a window 
on nonperturbative QCD in mesons with two very distinct mass scales. Moreover, the strong coupling $g_{D^*\!D\pi}$ one can extract from  the decay width is related 
to a putative universal coupling $\hat g$ between heavy-light mesons and a low-momentum pion in the heavy-meson chiral Lagrangian~\cite{Casalbuoni:1996pg,
El-Bennich:2021ldv,Braghin:2021qmu}. At leading order in the $1/m_D$ expansion this relation is $g_{D^*\!D\pi} = 2\sqrt{m_D^*m_D}\,  \hat g /f_\pi$.

These calculations were based on one-covariant models of the $D$ and $D^*$ wave functions and were therefore not Poincar\'e invariant, so that the momentum partition 
parameters had to be chosen according to some semi-classical criterium~\cite{El-Bennich:2010uqs}. This, of course, was not satisfactory and the motivation remained 
to compute the decay amplitude guided by Ref.~\cite{Jarecke:2002xd} which dealt with the decays $\rho\to \pi\pi$, $\phi \to \bar KK$ and $K^*\to K\pi$. In this work we close
this gap and compute the vector-meson decay with the complete Poincar\'e covariant structure of the BSA for the $D$, $D^*$ and pion. Along the way, we also obtain 
the strong couplings considered in Ref.~\cite{Jarecke:2002xd} which we update.

The remainder of this paper is composed of five sections: in Section~\ref{strongdecay} we explain the framework in which the strong couplings are calculated and 
define the decay kinematics; in Section~\ref{PSVsec} we describe the functional approach to QCD we use to calculate the quark propagators and BSA of the mesons 
within a given truncation scheme in Euclidean space; in Section~\ref{mesonsprop} the numerical method to solve the BSE is summarized and the meson's masses 
and weak decay constants are calculated. Finally, in Section~\ref{secresult} we present our results for the strong couplings and wrap up with final remarks in Section~\ref{conclude}.


\section{Strong Decay Amplitude\label{strongdecay}}

\begin{figure}[t!] 
\centering
  \includegraphics[scale=0.33,angle=0]{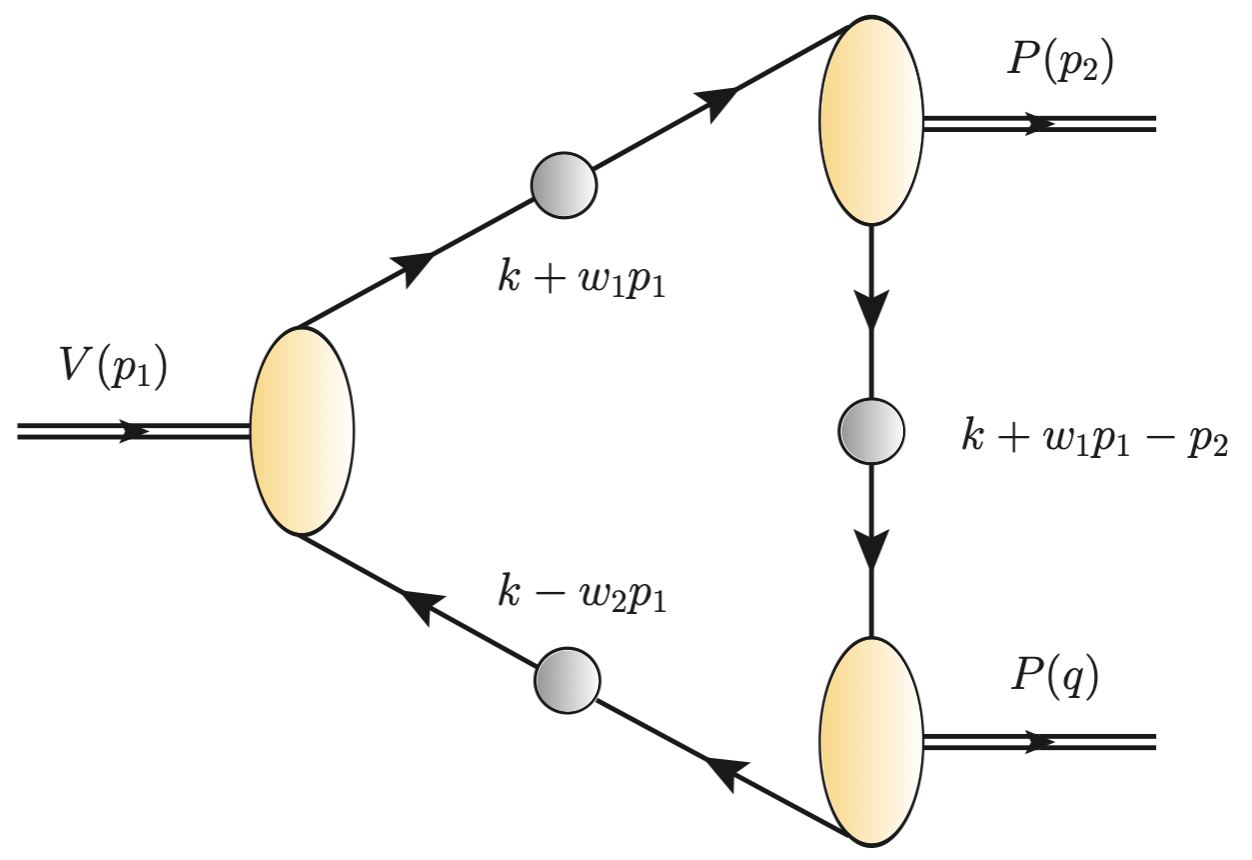} 
   \caption{Decay diagram depicting a generic strong decay $V\to PP$ in the impulse approximation of Eq.~\eqref{decayamp}. The shaded ovals represent vector ($V$) and
                 pseudoscalar ($P$) meson BSAs~\eqref{BSAdecomp}, while the dark-shaded circles symbolize dressed quark propagators~\eqref{DEsol} and the double-lined 
                 arrows describe the incoming vector-meson momentum and outgoing pseudoscalar-meson momenta.  }
 \label{Decaydiag} 
\end{figure}

In what follows, we limit ourselves to the \emph{impulse approximation} of the strong decays depicted in Fig.~\ref{Decaydiag}. As argued in Ref.~\cite{Jarecke:2002xd}, 
since the $\rho$ and $\phi$ appear as resonance poles in the timelike electromagnetic form factors of the pion and kaon, the pole residues are proportional to the 
respective coupling constants $g_{\rho\pi\pi}$  and $g_{\phi KK}$. Hence, if these form factors are obtained in impulse approximation, so will be the couplings. Now, the 
impulse approximation for the electromagnetic coupling to mesons conserves the current as long as the meson's BSA and quark-photon vertex are 
calculated in the ladder and the quark propagators in rainbow truncation, respectively, and the resulting electromagnetic form factors are in excellent agreement with 
experiment~\cite{Maris:2000sk,Chang:2013nia,Raya:2015gva,Ding:2018xwy}. 

In the time-like region, on the other hand, the ladder truncation of the BSE fails to produce the $\rho$ pole in $e^+e^- \to \gamma^* \to  \pi^+\pi^-$ and was amended to 
include effective pion degrees of freedom in the BSE scattering kernel~\cite{Williams:2018adr,Miramontes:2019mco,Miramontes:2021xgn}. Therefore, we expect that the 
impulse approximation for the strong decays of lighter mesons misses some of the relevant physics, in particular in the case of the $\rho$ meson whose decay width is
almost 20\% of its mass. Going beyond this approximation, not merely in the BSE kernel but also in the decay amplitude, is a technically and numerically challenging task. 
As the main aim here is to improve on earlier calculations of $D^*\to D\pi$, we deliberately ignore these corrections. 

The strong decay coupling for a process $V\to PP$ is defined as,
\begin{equation}
   \langle P(p_2) P(q) | V (p_1,\lambda) \rangle : = \, g_{V\!PP} \, \epsilon^\lambda \!\cdot q \ ,
\end{equation}
where the initial state is a vector meson with transverse polarization $\epsilon^\lambda_\mu$ and momentum $p_1^2=-m_V^2$, while the light(er) mesons have on-shell 
momenta $p_1^2 = -m_P^2$, $q^2 = -m_P^2$, with $q=p_1-p_2$, and can have different flavor content. The decay amplitude in impulse approximation can be expressed 
by the loop integral:
\begin{align} 
    g_{V\!PP} \; \epsilon^\lambda \!\cdot q  \, &  =  \int^{\Lambda}\! \frac{d^4 k}{(2\pi)^4}\, \mathrm{Tr_{CD}} \, \big [ \epsilon^\lambda\! \cdot \Gamma_V (k_V, p_1) 
    \nonumber \\
    S_f (k_1 ) & \bar \Gamma_P (k_P, -p_2) S_f (k_2) \bar \Gamma_P (k_P',-q ) S_f (k_3) \big ] \ .
\label{decayamp}    
\end{align}
In here, $\Gamma (k,P)$ are the BSAs of the mesons, $S(k_i)$ are the quarks propagators and the trace is over color and Dirac indices. Following the momentum flow 
in Fig.~\ref{Decaydiag}, the quark momenta are defined as, 
\begin{align}
   k_1  & = k + w_1 p_1 \, , \\ 
   k_2  & = k + w_1 p_1 - p_2 \, ,\\  
   k_3  & = k - w_2 p_1 \, . 
\end{align}
with the constraint $w_1 + w_2 = 1$ on the partition parameters due to momentum conservation. The relative BSA momenta are given by, 
\begin{align}
   k_V  & = k + \tfrac{1}{2} (w_1-w_2) p_1 \, ,  \label{relmomV} \\ 
   k_P  & = k + w_1 p_1 -  \tfrac{1}{2}\,  p_2 \, ,  \label{relmomP1}  \\  
   k_P'  & = k +  \tfrac{1}{2} (w_1-w_2) p_1 -  \tfrac{1}{2}\,  p_2    \label{relmomP2}  \, . 
\end{align}
Note that the relative momentum of the vector meson is only real if $w_1= w_2$, as in the meson's rest frame $p_1=(\mathbf{0}, i m_V)$ in the Euclidean-space formulation
we use. This will be discussed in more detail in Section~\ref{secresult}. 

We conclude this section by mentioning some definitions with respect to the charge when one of the final mesons is an isovector state. We follow Ref.~\cite{Bracco:2011pg}
and define the generic $D^*D\pi$ coupling as the one containing the neutral meson:
\begin{align}
  g_{D^*\!D \pi}  & = \, g_{D^{* \pm} D \mp \pi^0}=g_{D^{* 0} D^0 \pi^0} \nonumber \\
                          & =\, \tfrac{1}{\sqrt{2}} \, g_{D^{*-} D^0 \pi^{+}} = \tfrac{1}{\sqrt{2}} \, g_{D^{*+} D^0 \pi^{-}} \ .
\end{align}
Likewise, considering SU(3) flavor algebra one has, 
\begin{equation}
  g_{K^*\!K \pi}  =  \sqrt{3}\, g_{K^{*+} K^{+} \pi^0}=\sqrt{\tfrac{3}{2}} \,  g_{K^{*+} K^0 \pi^{+}} \ ,
\end{equation}
and moreover $g_{\rho\pi \pi} =  g_{\rho^0 \pi^+ \pi^-}$, $g_{\phi K K} =  g_{\phi K^+ K^-}$. 

In Section~\ref{PSVsec} we describe how the ingredients of the strong decay  amplitude~\eqref{decayamp}, namely the quark propagators and BSAs, are obtained 
from solving the quark-gap equation and the BSE.


\section{Pseudoscalar and Vector Meson Bound States\label{PSVsec}}


\subsection{Bethe-Salpeter Equation}

The relativistic initial and final bound states in the decay amplitude~\eqref{decayamp} are described by Poincar\'e covariant BSAs, $\Gamma^{fg}_{P}(k,P)$ and 
$\Gamma^{fg}_{V\nu}(k,P)$, which are the solutions of the homogeneous BSE in the $J^{PC} = 0^{-+}$  and $J^{PC} = 1^{--}$ channels, respectively~\cite{Bashir:2012fs}:
\begin{align}
  \Gamma^{fg}_{P}(k, P) =  \int^\Lambda \!  \frac{d^4q}{(2\pi)^4} \,  K^{fg}(k,q, P) \, \chi^{fg}_{P}  (q,P)  \ ,  
\label{BSE-pseudo}  \\
  \Gamma^{fg}_{V\nu}(k,P) =  \int^\Lambda \!  \frac{d^4q}{(2\pi)^4} \, K^{fg}(k,q, P) \, \chi^{fg}_{V\nu}  (q,P)  \ .
\label{BSE-vec}
\end{align}
In these BSEs, $k$ is the relative quark-antiquark momentum, $P$ is the meson momentum and $K^{fg}(k,q,P)$ is the fully amputated scattering kernel which sums 
up all possible quark-antiquark interactions. The Bethe-Salpeter wave functions, $\chi^{fg}_{P} (k,p)$ and $\chi^{fg}_{V\nu} (k,P)$, are obtained by attaching the quark 
propagators to the BSA,
\begin{align}
   \chi^{fg}_{P} (k,P) & = \ S_f(k_\eta)\, \Gamma^{fg}_{P}(k,P) S_g(k_{\bar\eta}) \ ,  \\
   \chi^{fg}_{V\nu} (k,P) & = \ S_f(k_\eta)\, \Gamma^{fg}_{V\nu}(k,P) S_g(k_{\bar\eta}) \ , 
\end{align}
with the shorthands, $k_{\eta} = k+\eta P$ and $k_{\bar\eta} = k-\bar\eta P$, that define momentum-partition parameters: $\eta + \bar\eta = 1$.

The BSA has the most general Poincar\'e covariant form that can be composed of the Dirac matrices and the relative and total momenta consistent with the quantum 
numbers $P$ and $C$ of a given meson,
\begin{equation}
   \Gamma_{M}^{fg} (k, P)=\sum_{i=1}^N \, T^i(k, P)\, \mathcal{F}_i^{fg}  \big (k, P, z_k \big ) \ ,
 \label{BSAdecomp}  
\end{equation}
where $ T^i(k, P) $ are Dirac covariants, $\mathcal{F}_i^{fg}$ are scalar Lorentz-invariant amplitudes and $z_k=k \cdot P /|k \| P|$ is an angle between $k$ and $P$.
In case of pseudoscalar mesons, we choose the usual $N=4$ covariants,
\begingroup
\addtolength{\jot}{0.2em}
\begin{align}
  T^1(k, P) & =i \gamma_5 \,  \\
  T^2(k, P) & =\gamma_5\, \gamma \cdot P  \ , \\
  T^3(k, P) & =\gamma_5\, \gamma \cdot k\,  k \cdot P \ , \\
  T^4(k, P) & =\gamma_5\, \sigma_{\mu \nu} \, k_\mu P_\nu \ ,
\end{align}
\endgroup
and for a vector meson $N=8$ covariant vector components are required:
\begingroup
\addtolength{\jot}{0.2em}
\begin{align}
T^1_{\nu}(k,P) & = \   i\gamma^T_\nu \ ,  \label{Tvec1}  \\
T^2_{\nu}(k,P) & = \  i \left [  3k^T_\nu \gamma\cdot k^T - \gamma^T_\nu \big ( k^T \big )^{\!2} \right ] \ , \\
T^3_{\nu }(k,P) &= \  i k\cdot P \, \gamma\cdot P\,  k^T_\nu \ ,  \\
T^4_{\nu}(k,P) & = \  i \left [\gamma^T_\nu \gamma\cdot P \, \gamma\cdot k^T + k^T_\nu \, \gamma \cdot P \right ] \ , \\
T^5_{\nu}(k,P)& = \ k^T_\nu \ , \\
T^6_{\nu}(k,P)& = \  k\cdot P  \left [ \gamma^T_\nu \gamma\cdot k^T -  \gamma\cdot k^T \,\gamma^T_\nu \right ] \ , \\
T^7_{\nu}(k,P)& = \ \gamma^T_\nu \gamma\cdot P - \gamma\cdot P\, \gamma^T_\nu -2\, T_\nu^8(k,P) \ , \\
T_{\nu}^8(k,P) & = \  \hat k^T_\nu \, \gamma\cdot \hat k^T \,  \gamma\cdot P \ .   \label{Tvec8}
\end{align}
\endgroup
The transverse projections are  $V^T_\nu = V_\nu -P_\nu(V\cdotp P)/P^2$ with $P\cdot V^T =0$ for any four-vector $V_\nu$ and $\hat k^T \cdot\hat k^T=1$.
Note that the $T^i_{\mu}(k,P)$ in Eqs.~\eqref{Tvec1} to \eqref{Tvec8} form an orthogonal basis~\cite{Maris:1999nt,Gao:2014bca} with respect to the Dirac trace.

In order to calculate the meson's weak decay constant, one has to normalize the meson's BSA. We do so with the derivative of the eigenvalue trajectory,
$\lambda (P^2)$, of the BSE~\cite{Nakanishi:1965zz,Nakanishi:1965zza}:
\begin{align}
  \left(\frac{\partial \ln \lambda }{\partial P^{2}}\right)^{\!-1} \!\! & = \  \operatorname{tr}_{\mathrm{CD}}  \int \frac{d^4k}{(2\pi)^4}  \, \bar{\Gamma}^{fg}_M (k ;-P) 
  \nonumber \\  
                       & \times  \ S_f  (k_{\eta } ) \Gamma^{fg}_M  (k ; P) S_g (k_{\bar \eta })  \, .
\label{nakanishinorm}                      
\end{align} 

With this we calculate the weak decay constant of the pseudoscalar meson defined by
\begin{equation}
   f_P  P_\mu   =  \langle 0 \, | \bar q_g \gamma_5 \gamma_\mu q_f |  P  (k,P) \rangle  \ ,
\end{equation}
which can be expressed by the integral:
\begin{equation}
\label{fdecay} 
  f_P  P_\mu = \frac{\mathcal{Z}_2 N_c }{\sqrt{2} }\int^\Lambda\!  \frac{d^4k}{(2\pi)^4} \,\operatorname{Tr}_\mathrm{D} 
     \left [ \gamma_5\gamma_\mu\,  \chi_P^{fg} (k,P) \right ] \, .
\end{equation}
Likewise, the weak decay constant of a vector meson is defined by the amplitude,
\begin{equation}
   f_V  m_V\,  \epsilon^\lambda_\mu   =  \langle 0\, | \bar q_g \gamma_\mu q_f |  V  (k,P,\lambda) \rangle  \ ,
\end{equation}
where $m_V$ is the vector-meson mass and $\epsilon^\lambda_\mu (P)$ is the polarization vector of the transverse vector meson of helicity $\lambda$ which 
satisfies $\epsilon^\lambda \cdot P = 0$  and is normalized as ${\epsilon^{\lambda}}^* \!\cdot \epsilon^\lambda = 3$. This can again be expressed 
by a loop integral:
\begin{equation}
 f_V  m_V  =   \frac{\mathcal{Z}_2 N_c}{3\sqrt{2}}  \int^{\Lambda}\!\! \frac{d^4k}{(2\pi)^4} \, \operatorname{Tr_D} \left [  \gamma_\mu \, \chi^{fg}_{V\mu}(k,P)  \right ] \ .
\label{vectordecay}
\end{equation}
In both expressions for the decay constants we define $\mathcal{Z}_2 (\mu,\Lambda) =\surd Z_2^f\surd Z_2^g$, as in Section~\ref{truncscheme}, and $N_c = 3$.


\subsection{Quark Gap Equation}

Amongst the Green functions that enter the BSE, whether in Eq.~\eqref{BSE-pseudo} or Eq.~\eqref{BSE-vec}, are the flavor-dependent dressed quark propagators 
described by Schwinger functions we obtain as solutions of the Dyson-Schwinger equation (DSE),
 \begin{align} 
    S^{-1}_f (p)   =  & \ Z_2^f \! \left (i\, \gamma \cdot  p + m^{\mathrm{bm}}_f \right )  \nonumber   \\
                    + \ Z_1^f & g^2 \!\! \int^\Lambda \! \frac{d^4k}{(2\pi)^4}  \, D^{ab}_{\mu\nu} (q) \frac{\lambda^a}{2} \gamma_\mu S_f(k) \, \Gamma^b_{\nu,f}  (k,p) \, ,
\label{QuarkDSE}
\end{align}
where $m^\textrm{bm}_f$ is the bare current-quark mass, $Z_1^f(\mu,\Lambda)$ and $Z_2^f(\mu,\Lambda)$ are the vertex and  wave-function renormalization 
constants at the renormalization point $\mu$, respectively. The integral in Eq.~\eqref{QuarkDSE} represents the self-energy of the quark and involves the dressed-quark 
propagator $S_f(k)$, the dressed-gluon propagator $D_{\mu\nu}(q)$ with momentum $q=k-p$ and the quark-gluon vertex, $\Gamma^a_\mu (k,p) = \frac{1}{2}\,\lambda^a 
\Gamma_\mu (k,p)$~\cite{Albino:2018ncl,Albino:2021rvj,El-Bennich:2022obe}, where the SU(3) color matrices $\lambda^a$  are in the fundamental  representation. 
The Poincar\'e-invariant regularization scale is $\Lambda \gg \mu$ and can be taken to infinity. The solution of the DSE can be cast in the most general covariant form as, 
\begin{align}
    S_f (p)  & =  \, -i \gamma \cdot p \, \sigma_{\rm v}^f ( p^2 ) + \sigma_{\rm s}^f ( p^2 ) \nonumber \\
                 & =   \, Z_f (p^2 ) / \left [ i \gamma \cdot p + M_f ( p^2 ) \right  ] \ .
\label{DEsol}                        
\end{align}
In this DSE, $Z_f(p^2)$ defines the wave function and $M_f (p^2)$ is the running mass of the quark. The scalar functions $\sigma_{\rm s}^f ( p^2 )$ 
and $\sigma_{\rm v}^f ( p^2 )$ thus depend on $Z_f (p^2)$ and $M_f (p^2)$. In a subtractive renormalization scheme the two renormalization conditions,
\begin{align}
   Z_f ( \mu^2)  & = \, 1 \  ,
\label{EQ:Amu_ren}   \\
   S^{-1}_f ( \mu^2 ) & =  \,  i \gamma\cdot p \ + m_f(\mu ) \ ,
 \label{massmu_ren}
\end{align}
are imposed, where $m_f(\mu )$ is the renormalized current-quark mass related to the bare mass by,
\begin{equation}
\label{mzeta} 
   Z_4^f  (\mu,\Lambda )\, m_f(\mu)  = Z_2^f  (\mu,\Lambda ) \, m_f^{\rm bm} (\Lambda) \  ,
\end{equation}
and $Z_4^f(\mu,\Lambda )$ is the renormalization constant that pertains to the mass term in the QCD Lagrangian.


\subsection{Truncation Scheme\label{truncscheme}}

The rainbow-ladder (RL) truncation of the integral equation~\eqref{QuarkDSE} and of the BSE kernel has proven to be a robust and successful symmetry-preserving 
approximation and allows for the description of light ground-state mesons in the isospin-nonzero pseudoscalar and vector channels. The RL truncation is 
realized by restricting the fully dressed quark gluon vertex to the perturbative vertex:  $\Gamma_{\nu,f} \to Z_2^f \gamma_\nu$. The DSE kernel then reduces 
to~\cite{Serna:2018dwk},
\begin{equation}
\hspace{-1.5mm}
   Z_1^f g^2  D_{\mu\nu} (q) \Gamma_{\nu, f} (k, p) = \big ( Z^f_2 \big)^{\!2} \mathcal{G}_f (q^2) D_{\mu\nu}^\mathrm{free} (q) \frac{\lambda^a }{2} \gamma_\nu \, ,
\label{RLtrunc}
\end{equation}
in which an Abelianized Ward identity is enforced that leads to $Z_1^f = Z_2^f$~\cite{Bashir:2012fs} and implies the omission of the three-gluon interaction in 
$\Gamma_{\mu}(k, p)$.  An additional factor $Z_2^f$ in Eq.~\eqref{RLtrunc} ensures multiplicative renormalizability of the DSE and therefore the mass function  
$M_f (p^2)$ is a renormalization-point invariant quantity~\cite{Bloch:2002eq}.

We work in Landau gauge in which the free gluon propagator is transverse,
\begin{equation}
    D_{\mu\nu}^\mathrm{free} (q) :=  \delta^{a b}\left(\delta_{\mu \nu}-\frac{q_{\mu} q_{\nu}}{q^{2}}\right ) \! \frac{1}{q^2} \ ,
 \label{freegluon}   
\end{equation}    
and introduce the flavor-dependent interaction,
\begin{equation}
    \frac{\mathcal{G}_f(q^2)}{q^2}  = \,  \mathcal{G}_f^\mathrm{ IR} (q^2) +  4\pi \tilde\alpha_\mathrm{PT} (q^2)  \ ,
\label{IR+UV}    
\end{equation}
where we deliberately absorb a factor $1/q^2$ from the gluon propagator~\eqref{freegluon}. The dressing function $\mathcal{G}_f (q^2)$ consists of a term that 
dominates in the infrared domain, $ |k| < \Lambda_\mathrm{QCD}$, and is suppressed at large momenta, and a second term that implements the regular continuation
of the perturbative QCD coupling and dominates large momenta. We use the model of Ref.~\cite{Qin:2011dd} given by,
\begin{align}
    \mathcal{G}_f^\mathrm{IR} (q^2)  & =   \frac{8\pi^2}{\omega^4_f}  D_f\,  e^{-q^2/\omega^2_f}    
\label{G-IR}      \\
  4\pi \tilde\alpha_\mathrm{PT}(q^2) & =  \frac{8\pi^2  \gamma_m\, \mathcal{E}(q^2)}{ \ln \left [  \tau +
                       \left (1 + q^2/\Lambda^2_\textrm{\tiny QCD} \right )^{\!2} \right ] }  \ , 
\label{G-PT}                         
\end{align}
in which $\gamma_m=12/(33-2N_f)$ is the anomalous mass dimension and $N_f $ is the active flavor number, $\Lambda_\textrm{\tiny QCD}=0.234$~GeV, 
$\tau=e^2-1$, $\mathcal{E}(q^2)=[1-\exp(-q^2/4m^2_t)]/q^2$ and $m_t=0.5$~GeV.  

The flavor dependence of the interaction $\mathcal{G}_f^\mathrm{IR} (q^2)$ was introduced in Refs.~\cite{Serna:2017nlr,Serna:2020txe,Chen:2019otg,Serna:2022yfp} to 
accommodate the strong flavor-symmetry breaking effects that led to complications in the calculation $D$- and $B$-meson properties~\cite{Rojas:2014aka,Mojica:2017tvh}; 
see Refs.~\cite{Serna:2020txe,Serna:2022yfp,Qin:2019oar} for the details of the implementation of the BSE kernels~\eqref{BSE-pseudo} and \eqref{BSE-vec} consistent 
with Eq.~\eqref{RLtrunc}. Suffice to say that herein we employ $\mathcal{G}_u (q^2)=\mathcal{G}_d (q^2) = \mathcal{G}_s (q^2)\neq \mathcal{G}_c (q^2)\neq \mathcal{G}_b (q^2)$.

\begin{table}[t!]
\renewcommand{\arraystretch}{1.3}
\setlength{\tabcolsep}{5pt}
\centering
\begin{tabular}{c|c|c|c|c|c}
\hline \hline
  Flavor  &  $m_f (19\,\mathrm{GeV})$  & $m_f (2\,\mathrm{GeV}) $  & $\omega_f $  & $\kappa$  &  $M^E_f$   \\ 
\hline
  $u,d$ &   0.0034 & 0.018 & 0.500 & 0.80 & 0.408 \\
   $s$ & 0.082 & 0.166 & 0.500 & 0.80 & 0.562   \\
   $c$ & 0.903 &1.272& 0.698  & 0.60 & 1.342   \\
   $b$ &   3.741& 4.370 & 0.640 & 0.56 & 4.259   \\
\hline \hline
\end{tabular}
\caption{Parameters of the interaction model in Eqs.~\eqref{G-IR} and \eqref{BSEflavor}:  $m_f (\mu)$,  $\omega_f$ and $\kappa=(\omega D_f)^{1/3}$ (in GeV). 
             $M^E_f$ is the Euclidean constituent quark mass: $M^E_f = \{p^2 | p^2=M^2(p^2) \} $.  }
\label{tab:parameters} 
\end{table}

The form of the quark-antiquark ladder kernel we therefore employ differs somewhat from the usual one in case of heavy-flavored mesons,
\begin{equation}
    K_{fg}(k,q,P)   = -  \mathcal{Z}_2^2 \,\frac{\mathcal{G}_{fg}  (l^2)}{l^2 }\, D_{\mu\nu}^\mathrm{free} (l)\, \frac{\lambda^a }{2} \gamma_\mu \frac{\lambda^a }{2} \gamma_\nu \ ,
  \label{RLkernel}    
\end{equation}
in which the relative momentum is $l=k-q$. In other words, in Eq.~\eqref{RLkernel} we combine the wave-function renormalization constants of both 
quarks, $ \mathcal{Z}_2 (\mu,\Lambda) =\surd{Z_2^f}\surd {Z_2^g}$, and use the averaged interaction,
\begin{equation}
    \frac{ \mathcal{G}_{fg}  (l^2) }{ l^2 }= \mathcal{G}_{fg}^\mathrm{ IR}(l^2) +  4\pi \tilde\alpha_\mathrm{PT}(l^2) \, ,
\end{equation}
which leads to a different treatment of the light and heavy quarks. The interaction in the low-momentum domain is given by the
Gaussian form,
\begin{equation}
 \mathcal{G}_{fg}^\mathrm{IR} (l^2)  =   \frac{8\pi^2}{(\omega_f\omega_g)^2} \sqrt{D_f\,D_g }\,e^{-l^2/(\omega_f\omega_g)} \, ,
\label{BSEflavor} 
\end{equation}
while $4\pi \tilde\alpha_\mathrm{PT}(q^2)$ is as in Eq.~\eqref{G-PT}. The parameters of this interaction model are listed in Table~\ref{tab:parameters}~\cite{Serna:2020txe}.

\begin{table*}[t!]
\renewcommand{\arraystretch}{1.3}
\setlength{\tabcolsep}{8pt}
\centering
\begin{tabular}{c|c|c|c||c|c|c}
\hline\hline
 & $m_M$ &$m^\mathrm{exp}_M$   &  $\epsilon_{m_M}$ [\%]    &   $f_M $  &  $f^\mathrm{exp/LQCD}_M $  &  $\epsilon_{f_M} $ [\%]  \\ \hline
 $\pi (u \bar d)$&0.140&0.138&1.45 &   $0.094$ & 0.092(1)  & 2.17  \\ 
 $K (u \bar s)$  &0.494 & 0.494 & 0.0 & $0.110$  &0.110(2)  &  0.0 \\ 
 $D (c \bar d) $  & $1.867 $ &1.864 &0.11 &  $0.144 $  &0.150 (0.5) & 4.00   \\ 
 \hline 
  $\rho (u\bar u)$&0.730&0.775&5.81 &0.145& 0.153(1) & 5.23\\
  $\phi(s\bar s)$  &1.070&1.019&5.20&0.187&0.168(1)  &11.31 \\
  $K^*(u\bar s)$  &0.883&0.896&1.45&0.163&0.159(1) &2.55 \\
  $D^*(c \bar u) $&2.021&2.009&0.60&  0.165 &0.158(6) &4.43   \\
\hline \hline
\end{tabular}
\caption{Masses and weak decay constants [in GeV] of ground-state pseudoscalar and vector mesons, $M=P,V$. The experimental mass values are taken from the Particle 
Data Group [PDG]~\cite{ParticleDataGroup:2022pth} and the leptonic decay constants for the $\rho$, $K^*$ and $\phi$ mesons are derived from their experimental decay 
width via $f_{V}^{2}=\frac{3 m_{V}}{4 \pi \alpha^{2} Q^{2}}\,  \Gamma_{V \rightarrow e^{+} e^{-}}$. The decay constant of the $D^*$ meson is a lattice-QCD prediction by the 
ETM collaboration~\cite{Lubicz:2017asp}. The relative deviations from experimental values are  given by $\epsilon_v  = | v^\textrm{exp.} - v^\textrm{th.} | / v^\textrm{exp.}$.}
\label{tab:psproperties}
\end{table*}

The only missing ingredient now is the quark propagator for complex momenta,  
\begin{equation} 
   S_f (q_\eta)  =   -i \gamma \cdot q_\eta \, \sigma_{\rm v}^f (q_\eta^2 ) +  \sigma_{\rm s}^f (q_\eta^2 )\, ,
\end{equation}   
and likewise for  $S_f (q_{\bar\eta } )$, as in Euclidean space the arguments $q_\eta^2$ and $q_{\bar\eta}^2$ define parabolas on the complex plane, 
\begin{eqnarray}
   q^2_\eta  & = &  q^2  - \eta^2 m^2_M  + 2 i \eta \, m_M  | q | z_q \  , \nonumber \\  
   q^2_{\bar\eta}  & = &  q^2  - {\bar\eta}^2 m^2_M  - 2 i \bar \eta \, m_M  | q | z_q \,  ,
\label{complexquark}   
\end{eqnarray}
where $z_q = q\cdot P /|q||P|$. We apply Cauchy's integral theorem as described, e.g., in Ref.~\cite{Krassnigg:2008bob} and obtain the solutions of the DSE
on the complex plane with the contour parametrization of the parabola defined in Ref.~\cite{Rojas:2014aka}; see Refs.~\cite{El-Bennich:2016qmb,Serna:2020txe} 
for graphic visualizations of $\sigma_{\rm s}^u (q_\eta^2 )$.


\section{Pseudoscalar and Vector Meson Properties\label{mesonsprop}}

Using the quark propagators on the complex momentum plane~\eqref{complexquark} and the BSE kernel~\eqref{RLkernel} in Eqs.~\eqref{BSE-pseudo} and \eqref{BSE-vec}, 
we treat the BSE as an eigenvalue problem~\cite{El-Bennich:2015kja,El-Bennich:2017brb}. For instance, in case of the vector mesons, the covariant decomposition in 
Eq.~\eqref{BSAdecomp} along with the orthogonality of the basis in Eqs.~\eqref{Tvec1} to \eqref{Tvec8} allow to recast the homogeneous BSE~\eqref{BSE-vec} with the 
kernel~\eqref{RLkernel} in a set of eight coupled-integral equations,
\begin{align}
\hspace*{-1.2mm}
    \mathcal{F}_i^{fg}  & \big (k, P, z_k \big ) =  -\tfrac{4}{3}\,  \mathcal{Z}_2^2 \int^\Lambda_q  \! \mathcal{G}_{fg} \big( l^2 \big)  D^{\mathrm{free}}_{\mu\nu}(l)
    \mathcal{F}_j^{fg}  (q,P, z_q)  \nonumber \\
  &  \times  \,\mathrm{Tr}_{\mathrm{D}}  \left [ T^i_\rho (k,P) \gamma_{\mu} S_f (q_\eta)  T^j_\rho (q,P) S_g (q_{\bar \eta})  \gamma_\nu \right ]  ,
   \label{coupledeq}
\end{align}
where the mnemonic shortcut for the integral represents the same integral with Poincar\'e-invariant cut-off as before. 
In solving this equation system numerically, we expand the scalar amplitudes in terms of Chebyshev moments,  $\mathcal{F}_{im}^{fg} ( k, P )$,
\begin{equation}
   \mathcal{F}_i^{fg} ( k, P,z_k ) = \sum_{m=0}^{\infty} \mathcal{F}_{im}^{fg} ( k, P )\, U_m (z_k) \ ,
\label{chebyshev}   
\end{equation}
which allows for a faster convergence. We consider $m=3$ Chebyshev polynomials, $U_m (z_p )$, of second kind. The eigenvalue problem for the vector 
$\bm{\mathcal{F}} := \{\mathcal{F}_i, \ldots  , \mathcal{F}_8 \} $ is then solved by means of Arnoldi factorization in the \texttt{ARPACK} library~\cite{Lehoucq1998}. 
Details of the practical implementation of \texttt{ARPACK} in a numerical treatment of the BSE are reviewed, for instance, in Refs.~\cite{Blank:2010bp,Rojas:2014aka}.

\begin{figure*}[t!] 
\centering
  \includegraphics[scale=0.6,angle=0]{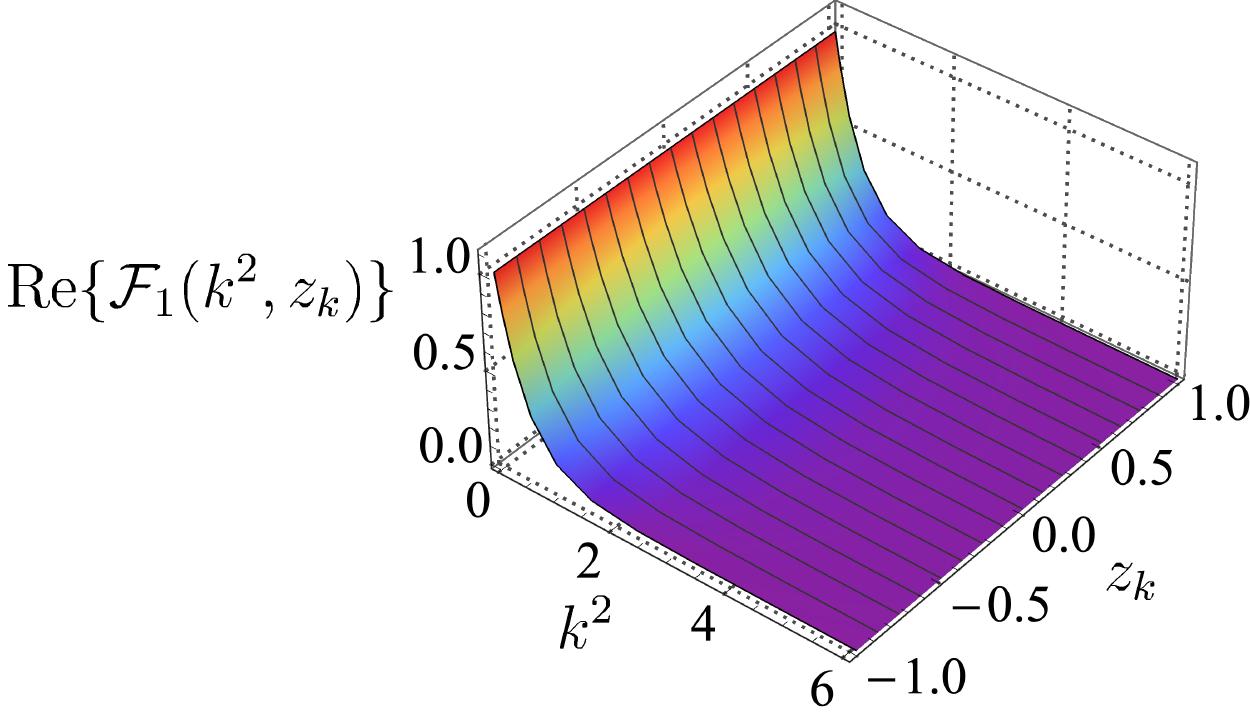}  \hspace*{6mm}
  \includegraphics[scale=0.6,angle=0]{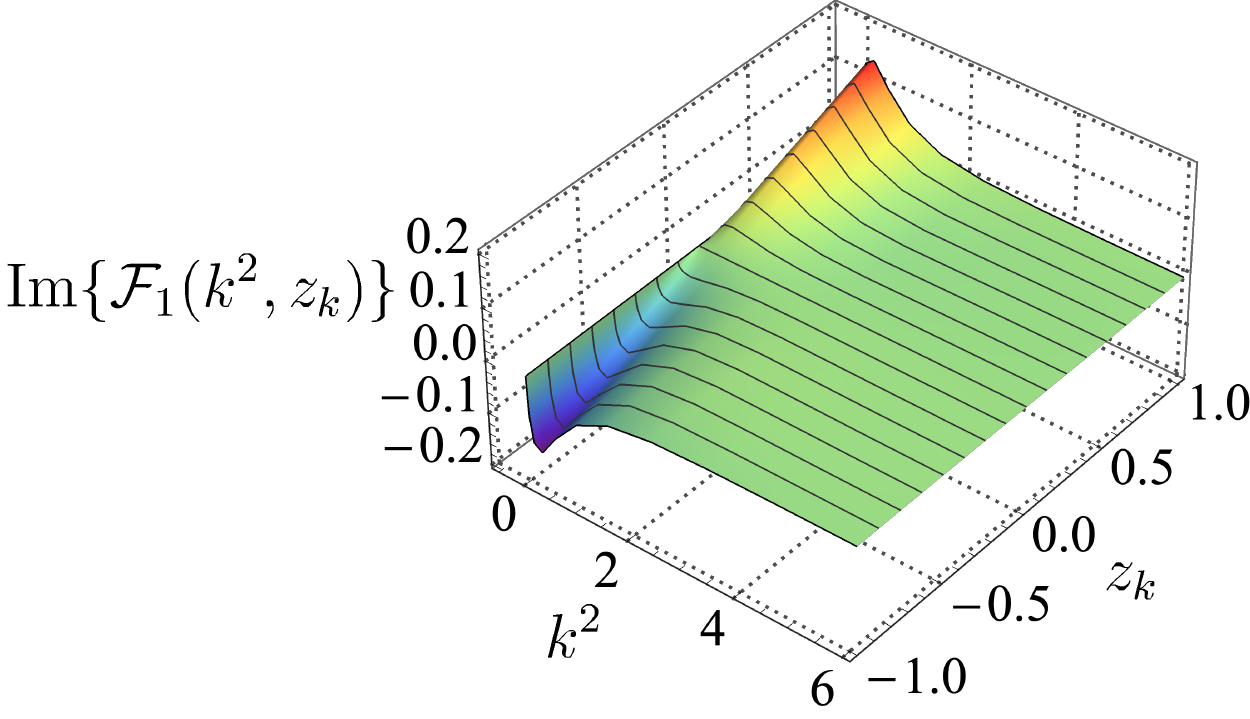} 
\vspace*{6mm} \\
   \includegraphics[scale=0.6,angle=0]{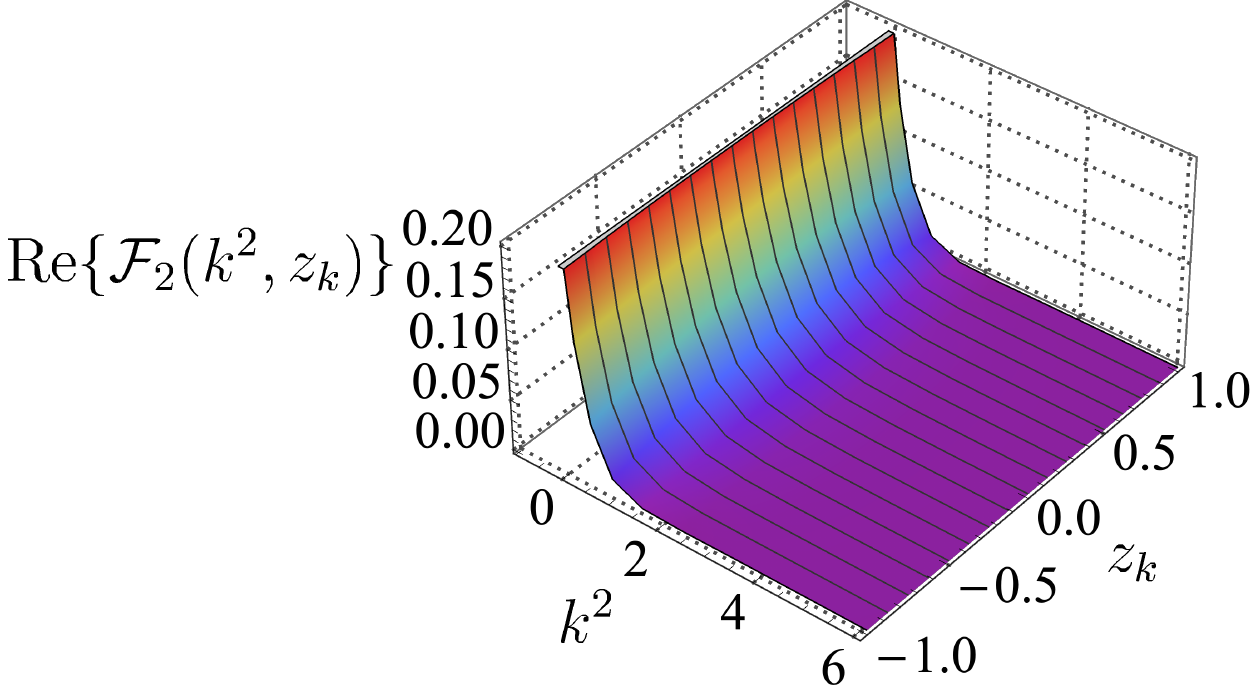}   \hspace*{6mm}
  \includegraphics[scale=0.6,angle=0]{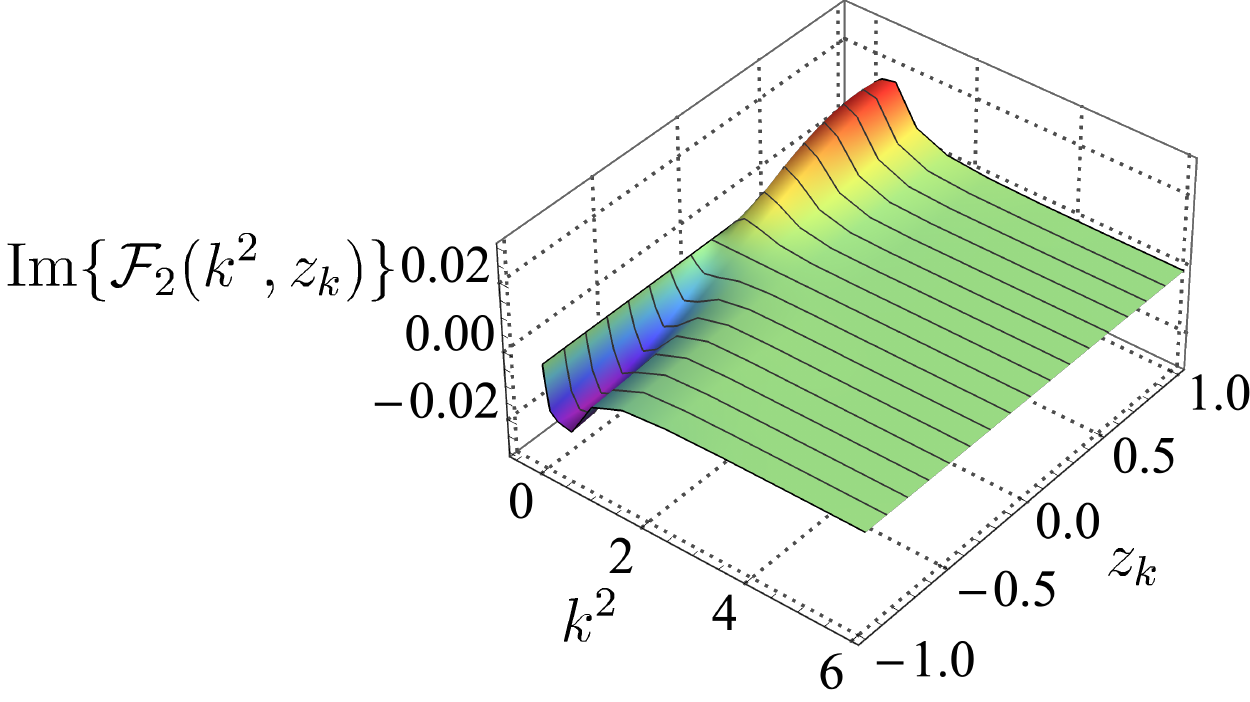} 
   \caption{\label{fig:Chebys}  Real and imaginary parts of the scalar amplitudes $\mathcal{F}_1(k,P)$ and $\mathcal{F}_2(k,P)$ of the  $D^*$ meson.} 
 \vspace*{3mm}  
\end{figure*}

\begin{figure*}[t!] 
\centering
  \includegraphics[scale=0.63,angle=0]{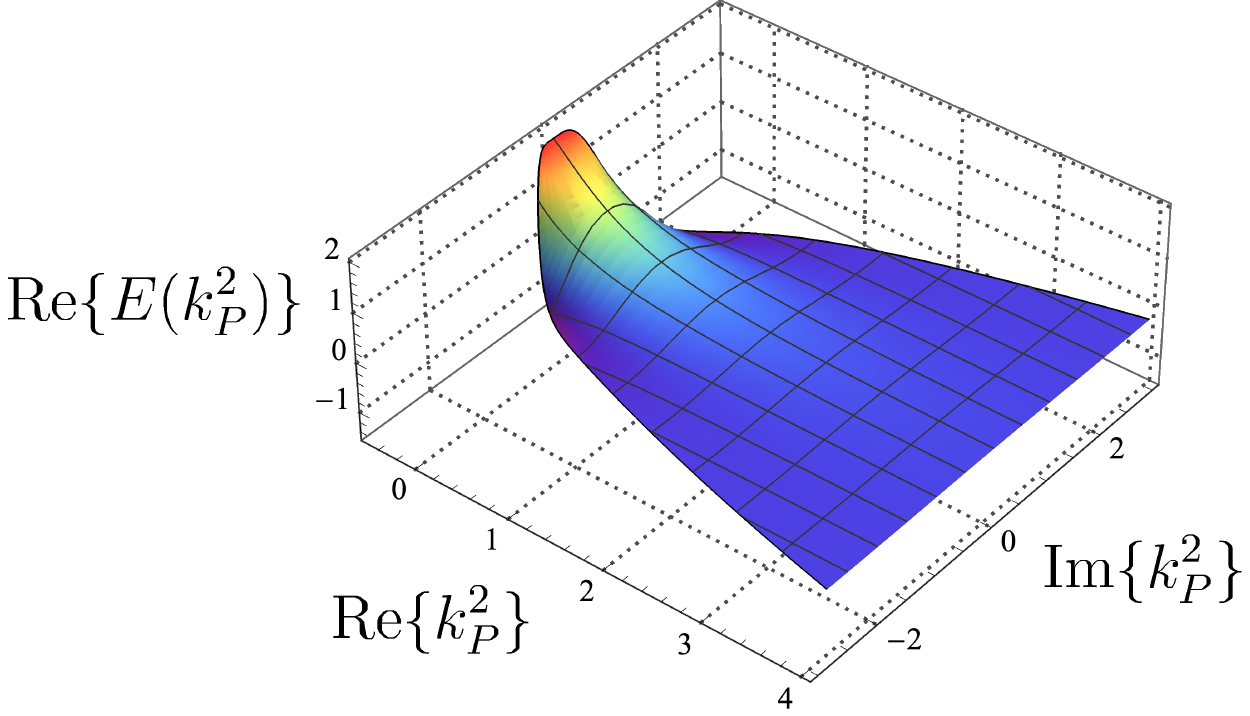}  \hspace*{6mm}
  \includegraphics[scale=0.63,angle=0]{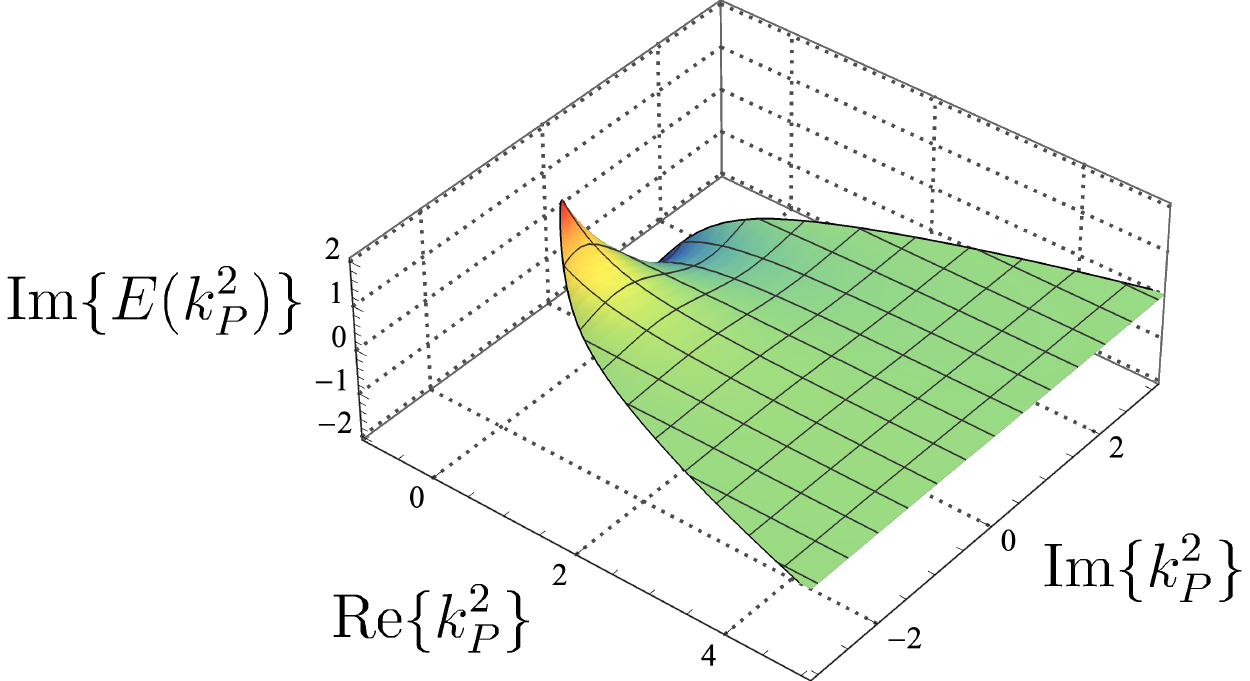} 
   \caption{\label{DmesonparamREIM}  Real and imaginary parts of the leading BSA of the $D$ meson analytically continued on the complex momentum plane
                            with the representation in Eq.~\eqref{BSAparKD}. Note that the amplitude is plotted on the parabola spanned by $k_P^2$~\eqref{relmomP1}  
                           whose vertex lies in the time-like region. On the real axis the BSA normalization is $E_D(k_p=0,p_2) =1$.}
\end{figure*}

The masses, $m_P$ and $m_V$, of the ground state  pseudoscalar and vector mesons are the solutions of the eigenvalue trajectory $\lambda (P^2 =-m^2_{P,V} )= 1$. 
They are listed  along with the leptonic decay constants in Table~\ref{tab:psproperties} and are in very good  agreement with experimental data, when available, or 
lattice-QCD results otherwise. As a byproduct we obtain the BSA of the mesons, which we illustrate with the real and imaginary parts of the dominant scalar functions 
of the $D^*$ meson in Fig.~\ref{fig:Chebys}.


\section{Results\label{secresult}}

As we work in Euclidean space, the relative momenta, $k_P$ and $k_P'$ in the decay, Eqs.~\eqref{relmomP1} and \eqref{relmomP2}, of the final pseudoscalar mesons 
and of the $D^*$ \eqref{relmomV} are complex. This is because in the center of mass of the initial vector meson its four-momentum is $p_1=(\mathbf{0}, i m_V)$ and thus 
$k_V= k +(w_1-w_2)p_1/2$ is only real if $w_1=w_2$. In case of the final-state mesons, the relative momentum is inevitably complex. In principle, due to the Poincar\'e 
invariance of the BSAs, our calculations are independent of the choice for the partition parameters. Practically, though, we are limited by numerical constraints 
as choosing $w_2 = 0.5$ in the case of $m_V = m_{D^*}$ implies probing the light-quark propagator at large time-like momenta, much larger than the light quark's 
mass. In this region, the solutions of the quark propagator on the complex momentum plane are characterized by branch cuts and/or complex-conjugate 
poles~\cite{El-Bennich:2016qmb,Eichmann:2021vnj} and a contour deformation is not trivial. 

Having in hand the numerical BSA for real momenta, we therefore parametrize it with a Nakanishi type of representation which allows for an analytical 
continuation of the $\mathcal{F}_i^{fg} ( k, P )$~\eqref{chebyshev} in the complex plane. In order to do so, we split the BSA in even and odd components,
\begin{equation}
\label{BSA-kaon-D}
   \mathcal{F}_i (k,P) = \mathcal{ F}^0_i (k,P) + k\cdot P\, \mathcal{F}^1_i (k,P) \, ,  
\end{equation}
in which $\mathcal{F}^{0,1}_i (k,P)$ are even under $k\cdot P \rightarrow   -k\cdot P$ and where we henceforth suppress the flavor indices $fg$. As discussed, 
for instance, in Ref.~\cite{Serna:2020txe}, $\mathcal{F}^1_i (k,P) \equiv 0$ for flavorless pseudoscalar mesons, such as the neutral pion, as they are eigenstates 
of the charge-conjugation operator defined as,
\begin{equation}
   \Gamma_M (k,P) \, \stackrel{C}{\longrightarrow} \  \bar{\Gamma}_M (k,P) :=C\, \Gamma^{T}_M (-k,P) C^{T} \ .
\end{equation}
The constraint that the covariant basis~\eqref{BSAdecomp} satisfies $\bar \Gamma_M (k,P)  = \lambda_c \Gamma_M (k,P)$ with $\lambda_c  = +1$ for pseudoscalar 
mesons and $\lambda_c  = -1$ for vector mesons, respectively, therefore imposes a definite parity of the scalar amplitudes $\mathcal{F}_i (k,P)$ under 
$k\cdot P \rightarrow  -k\cdot P$. For the neutral vector mesons,  $\rho$ and $\phi$, this implies that the $\mathcal{F}_i (k,P)$ are necessarily even and 
$\mathcal{F}^1_i (k,P) \equiv 0$ again. In case of the $K$, $K^*$, $D$ and $D^*$ mesons, which are not eigenstates of $C$, both amplitudes in Eq.~\eqref{BSA-kaon-D} 
must be considered.  

\begin{table*}[t!]
\centering 
\renewcommand{\arraystretch}{1.5}
\setlength{\tabcolsep}{6pt}

\begin{tabular}{c|c|c|c|c|c}
\hline\hline
    $g_{V\!PP}$       &  $(V,P)=(8,4)$  &  $(V,P)=(5,4)$  & $(V,P)=(1,1)$  &  Reference  & $\epsilon_{g_{V\!P\!P}} \ [\%] $   \\ \hline
    $g_{\rho\pi\pi}$   &  $5.13^{+0.24}_{-0.25}$   &  5.14  &  7.99   &  $5.94\pm 0.44$ &  13.6 \\
    $g_{\phi KK}$     &  $5.02^{+0.12}_{-0.22}$   &  5.03  & 10.12  &  $5.53\pm  0.31$ &  9.2 \\ 
    $g_{K^*\!K\pi}$   &  $5.10^{+0.22}_{-0.26}$   &  5.25  &   9.08  &   $5.47\pm 0.99$  & 6.8 \\ 
    $g_{D^*\!D\pi}$  &  $17.24^{+3.06}_{-2.30}$  & 16.41 &  37.22  & $ 17.9\pm 0.3 \pm1.90$  &  3.7  \\
\hline\hline
\end{tabular}
\caption{Strong couplings $g_{V\!PP}$ for the decay channels $\rho \to \pi\pi$, $\phi\to \bar KK$, $K^*\to K\pi$ and $D^*\to D\pi$. The pair $(V,P)$ denotes the number 
              of scalar amplitudes employed in the BSA of the vector and pseudoscalar mesons, respectively. The theoretical errors stem from the fit to Nakanishi representations
              of the BSA. The reference couplings are derived from the experimental decay widths~\cite{ParticleDataGroup:2022pth} via $\Gamma = g_{V\! P_1\! P_2}^2 k^3/ 
              6 \pi m_V^2$ with $k^2 = [m_V^2 - (m_{P_1} + m_{P_2} )^2 ] [m_V^2 - (m_{P_1} - m_{P_2})^2 ]/ 4m_V^2$. The experimental $D^*\!D\pi$ coupling was obtained
              by the CLEO collaboration~\cite{CLEO:2001sxb}. The relative deviations from experimental values are defined as $\epsilon_{g_{V\!P\!P}}   = | g_{V\!PP}^\textrm{exp.} 
              - g_{V\!PP}^\textrm{th.} | / g_{V\!PP}^\textrm{exp.}$.}  
\label{couplingtable}
\end{table*} 

When the relative momentum of the meson is complex,  we choose the following analytic representation of the scalar amplitudes for $l=0,1$, 
\begin{equation}
  \mathcal{F}_i^l (k,P) = \sum_{j=1}^{N} \int_{-1}^{1} \! d\alpha\,  \rho_j (\alpha)\, \frac{U_j \Lambda^{2n_j}}{\Delta^{n_j} (k,\alpha,\Lambda) }  
  \, ,
 \label{BSAparKD}
\end{equation}
where $\Delta=k^{2}+\alpha k \cdot P+\Lambda^{2}$ and the spectral density $\rho_j (\alpha)$ is given by,
\begin{equation}
   \rho_j  (\alpha) = \tfrac{1}{2} \! \left ( C^{\nicefrac{1}{2}}_{0}(\alpha)+\sigma_j C^{\nicefrac{1}{2}}_{2}(\alpha)  \right ) \, .
 \label{rhropion}   
\end{equation}
$C^{\nicefrac{1}{2}}_{0}(\alpha)$ and $C^{\nicefrac{1}{2}}_{2}(\alpha)$ are Gegenbauer polynomials of order $1/2$. The parameters $U_j$, $\Lambda$, $n_j$ and $\sigma_j$ 
are listed in Tables~\ref{table:1}, \ref{table:2} and \ref{table:3} of Appendix~\ref{tablesofparam} for the pion, kaon, $D$ and $D^*$ mesons. We use \emph{real} numerical BSA 
solutions for the remaining mesons considered in this work, namely the $\rho$, $\phi$ and $K^*$. This approach, frequently employed in calculations of distribution 
amplitudes~\cite{Shi:2014uwa,Serna:2020txe}, differs from the method in Ref.~\cite{Jarecke:2002xd} based on a 2nd order Taylor expansion of  $\mathcal{F}_i (k,P)$ 
about the closest point on the positive real axis to the complex-valued relative momenta $k$. We checked that the BSA parameterization of Eq.~\eqref{BSAparKD} produces 
the correct weak decay constants of the pseudoscalar and vector mesons. An illustration of the analytic continuation of the $D$-meson's dominante amplitude with the 
parametrization of Eq.~\eqref{BSAparKD} is given in Fig.~\ref{DmesonparamREIM}.

With these technical considerations taken into account, we can take the trace and calculate the loop integral in Eq.~\eqref{decayamp} for different initial and final states 
in the strong two-body decay of a vector meson. The couplings, $g_{V\!PP}$, we obtain are listed in Table~\ref{couplingtable}, where the errors are due to the fit uncertainties 
of the Nakanishi representations. We remind that this error would increase if we included a systematic error of the ladder truncation, as explored in Ref.~\cite{Serna:2020txe} 
for example. More precisely, it is known that some typical observables, such as the pion and kaon masses and weak decay constants, are insensitive to a range
$\omega_i \pm \Delta \omega_i$, $i=u,d,s$, of the interaction parameter in Eq.~\eqref{G-IR}. Varying $\omega_i$ alters the BSA of the mesons and this can 
add to the uncertainty in the strong $V\to PP$ decay amplitude. However, modifying $\omega_{u,d}$ in the $D$ and $D^*$ mesons is a numerically delicate matter,
as no solution of the BSA is found for the uncertainties $\pm\Delta \omega_{u,d}$. We therefore abstain from including this source of error. 

\begin{figure}[b!] 
\centering
  \includegraphics[scale=0.36,angle=0]{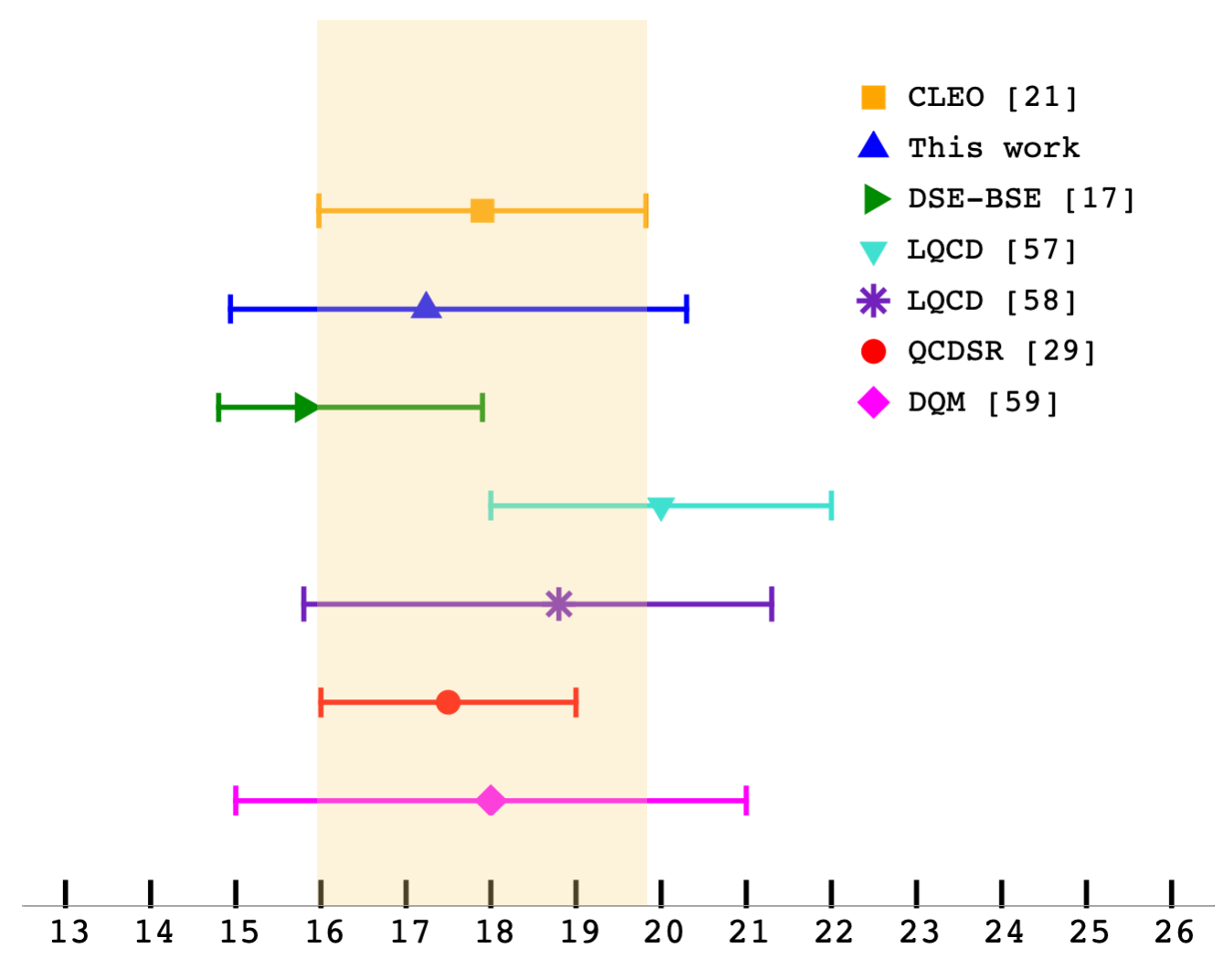} 
   \caption{Comparison of theoretical values for $g_{D^*\!D\pi}$ with the experimental coupling extracted from the $D^*$ decay width by the CLEO 
                 collaboration~\cite{CLEO:2001sxb} (shaded band). The couplings are taken from Ref.~\cite{El-Bennich:2010uqs} (DSE-BSE), Refs.~\cite{Becirevic:2009xp} 
                 and~\cite{Abada:2002xe} (LQCD), Ref.~\cite{Bracco:2011pg} (QCDSR) and Ref.~\cite{Melikhov:2001zv} (DQM). }
 \label{Dstarcomptable} 
\end{figure}

We consider the cases of limiting the BSA to the dominant covariant in Eq.~\eqref{BSAdecomp}, namely $\gamma_5$ and $\gamma_\mu$ for
the pseudoscalar and vector mesons, respectively. Using merely the dominant amplitude, $g_{\rho\pi\pi}$ is 56\% larger than the value obtained with the complete
BSA. We find a difference of 102\% for the $g_{\phi KK}$ coupling which increases to 116\% for $g_{D^*\!D\pi}$. Clearly, this leading approximation is not adequate
even for couplings that involve only light quarks. On the other hand, including the next four leading covariants of the vector meson's BSA, the dominant physics in 
the impulse approximation is captured and the couplings are within 1--5\% of the values obtained with $(V,P)=(8,4)$. 

Our values for the strong couplings mostly agree with reference values, except for the  $\rho\to \pi\pi$ coupling which is found to be the 13.6\% smaller than the 
experimental coupling. As mentioned in Section~\ref{strongdecay}, this is expected given our limitation to the impulse and ladder approximation which omits intermediate 
$\pi\pi$, $\bar KK$ and $K\pi$ channels. Including explicit two-pion exchange in the BSE kernel, the decay width of the $\rho$-meson can be determined from the 
imaginary part of the resonance pole from which one deduces a coupling constant $g_{\rho\pi\pi} = 5.7$~\cite{Williams:2018adr}. The widths of the $\phi$, 
$K^*$ and $D^*$, on the other hand, are much smaller than their masses and our approximation is more accurate, though we still notice a deviation of 9.2\% for $g_{\phi KK}$. 

As our focus is on the $D^*\to D\pi$ decay, we also compare our result with couplings obtained with lattice QCD (LQCD), QCD sum rules (QCDSR) and a dispersion relation 
quark model~(DQM) in Fig.~\ref{Dstarcomptable}. Our calculation is a significant improvement on earlier work~\cite{El-Bennich:2010uqs,El-Bennich:2011tme,
El-Bennich:2012hom,El-Bennich:2016bno} which also considered the impulse approximation but employed model wave functions for the mesons, based on the dominant 
covariant term of the BSA, and a simplified, constant-mass propagator for the charm quark. This, as we noted in Table~\ref{couplingtable}, has detrimental effects 
on translational invariance and the couplings depend on a \emph{suitable} choice of the partition parameters $w_1$ and $w_2$, see the discussion in 
Ref.~\cite{El-Bennich:2010uqs}. Since our calculation is fully Poincar\'e covariant, our decay amplitudes are independent of the momentum distribution, as we 
verified with variations of $w_1$ and $w_2$ up to a critical limit where  we encounter singularities in the quark propagators on the complex plane.


\section{Conclusion\label{conclude}}

We revisited the strong decays of vector mesons into two pseudoscalar mesons within the framework of the DSE and BSE, having in mind the particular decay 
$D^*\to D\pi$. As we argued, these decays are the simplest hadronic observables beyond the meson's masses, weak decay constants and electromagnetic form factors, 
and thus provide additional information about the dynamics of QCD in the nonperturbative regime. In particular, the vector-meson decay to a pair of pseudoscalars
proceeds via a $P$-wave interaction and therefore involves the BSA differently than the meson's weak decay constants. The strong $D^*$ decay is then even more
interesting, as it probes nonperturbative QCD simultaneously at two distinct scales, namely the light- and charm-quark masses. 

We limited ourselves to the impulse approximation for the aforementioned reason: our BSE kernel in ladder truncation is too simple to include $\pi\pi$, $\bar KK$ and
$K\pi$ channels in these decays and this is most likely the largest source of error in our calculation of the $\rho \to \pi \pi$ coupling. Nonetheless, this calculation represents 
an important theoretical and numerical improvement over the simpler approaches in Ref~\cite{El-Bennich:2010uqs,El-Bennich:2011tme,El-Bennich:2012hom,
El-Bennich:2016bno}, as the full Poncar\'e invariant BSA structure of all mesons is included and the quark propagators are calculated on the complex 
momentum plane for all flavors. This present calculation can also serve as a guidance to reevaluate off-shell space-like couplings between the $\rho$ and $D$- and 
$D^*$ mesons in Ref.~\cite{El-Bennich:2016bno} without resorting to model wave functions.

Our final value for $g_{D^*\!D\pi}$ is 3.7\% lower than that extracted from the experimental decay width and well within the experimental errors. It corresponds to 
a universal coupling in a chiral heavy meson Lagrangian which at leading order in the heavy-mass expansion is given by:
\begin{equation}
   \hat g =  \frac{g_{D^*\!D\pi} }{2\sqrt{m_D^*m_D} } \, f_\pi =  0.58^{+0.10}_{-0.08} \ .
\end{equation}
A consensus seems to be growing that the most recent theoretical couplings are in good agreement with the CLEO value~\cite{CLEO:2001sxb} extracted from the 
$D^*$ decay width. Future improvements ought to consider strong $\pi\pi$ interactions, likely along the lines presented in 
Refs.~\cite{Williams:2018adr,Miramontes:2019mco,Miramontes:2021xgn}, in the BSE kernels and to go beyond the impulse approximation.


\acknowledgements

B.E., F.E.S. and R.C.S. participate in the Brazilian network project \emph{INCT-F\'isica Nuclear e Aplica\c{c}\~oes\/}, no.~464898/2014-5. This work was supported by 
the S\~ao Paulo Research Foundation (FAPESP), grant no.~2018/20218-4, and by the National Council for Scientific and Technological Development (CNPq), 
grant no.~428003/2018-4.  F.E.S. is a CAPES-PNPD postdoctoral fellow financed by grant no.~88882.314890/2013-01.  We appreciated helpful communication
with Peter Tandy.  \\ \vspace*{5mm}


\onecolumngrid
\appendix
\vspace*{1cm}

\section{Parameters of Bethe-Salpeter Amplitude Representation\label{tablesofparam} }

\begin{table*}[h!]
\centering
\renewcommand{\arraystretch}{1.3}
\setlength{\tabcolsep}{6pt}
\begin{tabular}{c | r | r r r | r r r | r r r } 
 \hline\hline
 $\mathcal{F}_i^l$ & {$\Lambda$}  & {$U_{1}$} & {$U_{2}$}& {$U_{3}$} &{$\sigma_{1}$} & {$\sigma_{2}$}& {$\sigma_{3}$} &{$n_{1}$} &{$n_{2}$} &{$n_{3}$}\\ [0.5ex] 
 \hline
  $E_{\pi}^0$  & 1.280  & 2.558 & $-1.559$ & 0.0 & 1.810 & 1.548 & 0.0 & 4 & 5 &0  \\ 
  $F_{\pi}^0$  & 1.150  & 1.838 & $-0.948$ & $-0.381$ & $-2.679$ & $-2.547$ & $-5.107$ & 4 & 5 & 3   \\ 
  $G_{\pi}^0$  & 1.106  & 2.402 & $-1.950$ & 0.0 & $-0.4590$ & $-0.474$ & 0.0 & 6 & 7 & 0   \\  
  $H_{\pi}^0$  & 1.056  & 1.253 &  $-0.857$ & $-0.140$ & $-0.696$ & $-0.634$ & $-2.663$ & 5 & 6 & 3  \\ 
 \hline\hline
\end{tabular}
\caption{Parameters of the BSA representation in Eq.~\eqref{BSAparKD} for the pion. In the isospin limit, $m_u = m_d$, only $\mathcal{F}_{i}^0$ contributes which we  
              fit to the sum of the 0th and 2nd Chebyshev moments.}
\label{table:1}
\end{table*}    


\begin{table*}[h!]
\centering
\renewcommand{\arraystretch}{1.2}
\setlength{\tabcolsep}{7pt}
\begin{tabular}{c | r | r r r | r r r | r r r } 
 \hline \hline
 $\mathcal{F}_i^l$ & {$\Lambda$} & {$U_{1}$} & {$U_{2}$}& {$U_{3}$} &{$\sigma_{1}$} & {$\sigma_{2}$}& {$\sigma_{3}$} &{$n_{1}$} &{$n_{2}$} &{$n_{3}$}\\ [0.5ex] 
 \hline
   $E^0$  & 1.557  & 2.590 & $-1.590$ &  0.0  & 1.342 & 0.891 &  0.0 & 5 & 6 &0   \\   
   $E^1$  & 1.495  & $-0.810$ & 3.769 & $-2.251$ & $-1.039$ & $-0.680$ & $-0.680$ & 5 & 6 & 7  \\ 
   $F^0$  & 1.514  & 2.756  & $-3.558$  & 1.220  &  $-0.527$ &  $-0.173$  &  0.323 & 7 & 8 &  9  \\   
   $F^1$  & 1.604  & 3.150 & $-5.480$  &  2.537  &  $-1.074$ & $-0.881$ & $-0.718$ &10 & 11 &12  \\ 
   $G^0$  & 1.631  & $-0.613$ &1.139 & $-0.522$ & 0.595 & 2.134 & 2.651 & 8 & 10 & 12    \\ 
   $G^1$  & 1.229  & 2.949 & $-4.542$ & 1.880 & $-0.558$ & $-0.747$ &  $-0.920$ & 8 & 10 & 12   \\   
   $H^0$  & 1.727  & $-0.276$ & 1.031 &  $-0.594$ & 1.722 &  0.661 &  0.303 & 8 & 10 & 12   \\ 
   $H^1$  & 1.443  & $-1.564$ & 4.434 & $-2.663$ & $-0.8104$ & $-0.905$ & $-0.938$ & 8 & 9 & 10  \\ 
 \hline \hline
\end{tabular}
\caption{Parameters of the BSA representation in Eq.~\eqref{BSAparKD} for the kaon. Both, the $\mathcal{F}_{i}^0$ and $\mathcal{F}_{i}^1$ amplitudes contribute, 
              as the kaon is not an eigenstate of charge conjugation. In the fit we include the sum of the 1st and 3rd Chebyshev moments in the odd component of the BSA. }
\label{table:2}
\end{table*}


\begin{table*}[h!]
\centering
\renewcommand{\arraystretch}{1.2}
\setlength{\tabcolsep}{7pt}
\begin{tabular}{c | r | r r r | r r r | r r r } 
 \hline\hline
 $\mathcal{F}_i^l$ & {$\Lambda$} & {$U_{1}$} & {$U_{2}$}& {$U_{3}$} &{$\sigma_{1}$} & {$\sigma_{2}$}& {$\sigma_{3}$} &{$n_{1}$} &{$n_{2}$} &{$n_{3}$}   \\
 \hline
   $E^0$   & 1.750  &  2.078  & $-1.077$ & 0.0 & $-1.274$ & $-1.126$ & 0.0 & 5 & 6 & 0  \\  
   $E^1$   & 2.146  & $-0.207$ & 0.209 & 0.0 & $-1.115$ & $-1.115$ & 0.0 & 6 & 9 & 0  \\ 
   $F^0$   & 2.222  & 0.060  &  0.155 & 0.0 &  $-0.934$ & $-2.211$  &  0.0 & 6 & 9 & 0  \\   
   $F^1$   & 2.583  & 0.003  & $-0.043$ & 0.0 &  $-1.372$ & $-1.757$ & 0.0 & 6 & 9 & 0  \\ 
   $G^0$  & 1.543  & $-1.596$ & 2.008 & $-0.427$ & $-1.439$ & $-1.596$ & $-1.795$ & 6 & 7 & 10  \\ 
   $G^1$  & 1.423  & 0.197 & 0.201 &  $-0.233$ & $-1.289$ & $-2.657$  &  $-2.097$ & 5 & 6 & 9   \\   
   $H^0$  & 1.711  & $-0.249$ & 0.868 & $-0.530$ & $-1.535$ & $-1.618$ & $-1.621$ & 8 & 9 & 10  \\ 
   $H^1$  & 1.155  & $-0.481$ & 0.937 & $-0.470$ & $-1.917$ & $-2.011$ & $-2.086$ & 6 & 7 & 8  \\ 
 \hline \hline
\end{tabular}
\caption{Parameters of the BSA representation in Eq.~\eqref{BSAparKD} for the $D$ meson. Even and odd components of the BSA are in terms of Chebyshev moments
             as described in Tables~\ref{table:1} and \ref{table:2}. }
\label{table:3}
\end{table*} 

 \twocolumngrid


\begin{table*}[t!]
\centering
\renewcommand{\arraystretch}{1.3}
\setlength{\tabcolsep}{8pt}
\begin{tabular}{c | r | r r r | r r r | r r r  } 
 \hline\hline
 $\mathcal{F}_i^l$ & {$\Lambda$} & {$U_{1}$} & {$U_{2}$}& {$U_{3}$} &{$\sigma_{1}$} & {$\sigma_{2}$}& {$\sigma_{3}$} &{$n_{1}$} &{$n_{2}$} &{$n_{3}$}  \\ 
 \hline
    $\mathcal{F}_1^0$  & 1.942  & 3.510 & $-2.509$ & 0.0 & $-1.666$  & $-1.851$ & 0.0 & 7& 8 & 0  \\  
    $\mathcal{F}_1^1$  & 0.998  & 2.441 & $-1.533$ & $-0.680$ & $-1.856$ & $-2.027$ & $-1.167$ & 4 & 5 & 3  \\ \
    $\mathcal{F}_2^0$  & 1.107 & 0.440  &-0.216 & 0.0 & $-1.656$ & $-1.312$ & 0.0 & 4 & 6 & 0  \\  
    $\mathcal{F}_2^1$  & 0.902  & 2.493  &  $-3.754$ & 1.278 & $-2.500$ & $-2.522$ & $-2.550$ & 7 & 8 & 10  \\ \
    $\mathcal{F}_3^0$  & 1.445  & $-1.536$ & 1.372 & 0.0 & $-1.838$ & $-1.831$ & 0.0 & 7 & 8 & 0  \\ 
    $\mathcal{F}_3^1$  & 1.005  & 0.099 &  $-0.032$ & 0.0 & $-1.030$ & $-4.365$ & 0.0 & 3 & 9 & 0  \\ 
    $\mathcal{F}_4^0$  & 1.411 & $-1.235$ & 0.837  & 0.0 & $-1.844$ & $-1.833$ & 0.0 & 5 & 6 & 0   \\ 
    $\mathcal{F}_4^1$  & 1.554  &1.192 & $-2.367$ & 1.181 & $-1.949$ & $-2.040$ & $-2.10$ & 7 & 8 & 9  \\ 
    $\mathcal{F}_5^0$  & 1.895 & $-4.279$ & 2.958 & 0.0  & $-1.775$ & $-1.787$ & 0.0 & 7 & 8 & 0  \\ 
    $\mathcal{F}_5^1$  & 1.730 & 2.657 & $-4.797$ & 2.212 & $-1.831$ & $-1.932$ & $-2.00$ & 7 & 8 & 9  \\ 
    $\mathcal{F}_6^0$  & 1.384 & 0.815 & $-0.639$ & 0.0  & $-1.916$ & $-1.942$ & 0.0 & 6 & 8 & 0  \\  
    $\mathcal{F}_6^1$  & 1.250 & $-3.728$ & 2.097 & 0.0  & $-2.229$ & $-3.453$ & 0.0 & 7 & 12 &  0  \\ 
    $\mathcal{F}_7^0$  & 1.183 & $-0.557$ & 0.971 & $-0.412$ & $-0.711$ & $-1.517$ & $-1.927$ & 4 & 5 & 6    \\ 
    $\mathcal{F}_7^1$  & 1.091 & $-0.190$ & 0.076 & 0.080 & $-1.453$ & $-2.005$ & $-0.520$ & 4 & 8 & 3  \\ 
    $\mathcal{F}_8^0$  & 1.316 & 0.601 & $-0.746$ & 0.147  & $-1.917$ & $-2.011$ & $-2.086$ & 4 & 5 & 8    \\ 
    $\mathcal{F}_8^1$  & 0.909 & 0.073 & $-0.647$ & 0.501 & $-2.127$ & $-2.312$ & $-2.331$  & 4 & 5 & 6    \\  
 \hline \hline 
\end{tabular}
\caption{Parameters of the BSA representation in Eq.~\eqref{BSAparKD} for the $D^*$ meson. Even and odd components of the BSA are in terms of Chebyshev moments
             as described in Tables~\ref{table:1} and \ref{table:2}. }
\label{table:3}
\end{table*}



\end{document}